\documentclass[aps,pre,twocolumn]{revtex4-1}
\usepackage{amssymb}
\usepackage{amsmath,bm}
\usepackage{graphicx,color}
\usepackage{bbm}
\usepackage{multirow}
\newcommand{\B}[1]{{\bm{#1}}}

\usepackage{hyperref}

\begin{document}
\title{Fatigue and failure of a polymer chain under tension}
\author{Harish Charan$^1$, Alex Hansen$^{2,3}$,
	H.G.E. Hentschel$^{1,4}$ and Itamar Procaccia$^{1,5}$}
\affiliation{$^1$Dept. of Chemical Physics, The Weizmann Institute of
	Science, Rehovot 76100, Israel\\ $^2$ Dept. of Physics, Norwegian University of Science and Technology, Trondheim, Norway\\
	$^3$ Beijing Computational Sciences Research Center, CSRC, 10 East Xibeiwang Rd., Haidian District, Beijing 100193, China\\
$^4$ Dept. of Physics, Emory University, Atlanta Ga. 30322, USA\\$^5$  Center for OPTical IMagery Analysis and Learning, Northwestern Polytechnical University, Xi'an, 710072 China. }

\begin{abstract}
	The rupture of a polymer chain maintained at temperature $T$ under fixed tension is prototypical to a wide array of systems
	failing under constant external strain and random perturbations. 
	Past research focused on analytic and numerical studies of the mean rate of collapse of such a chain. Surprisingly, an analytic calculation 
	of the probability distribution function (PDF) of collapse rates appears to be lacking. Since rare events of rapid collapse can be important and even catastrophic, we present here a theory of this distribution, with a stress on its tail of fast rates. We show that the tail of the PDF is a power law
	with a {\em universal} exponent that is theoretically determined. Extensive numerics validate the offered theory. Lessons pertaining to other problems of the same
	type are drawn. 
\end{abstract}
\maketitle

{\bf Introduction}: The probability of rare events in which materials, devices or structures fail catastrophically, even when they are expected to remain whole for long {\em average} times,
is a subject of great interest in material physics, in engineering and in environmental sciences, cf. \cite{96SKKV,96Sch} and references therein. The development of techniques and ideas that allow the computation
of the probability of rare events is of obvious necessity. In this Letter, we discuss a fundamental problem of this type, and maybe one of the simplest, i.e. {\em a polymer in a thermal bath}, see \cite{01Put} and references therein. 
We consider a one dimensional chain of $N_b+1$ particles, or beads, interacting with their nearest-neighbors with a given potential. The chain is anchored at one end and pulled from the other end using a constant force $f$.
The chain is maintained at temperature $T$, and the question is how long will the chain persist until ``fatigue" will result in its breaking at time $\tau$. The breaking
time $\tau$ is a stochastic variable, since the dynamics at any given temperature induces random fluctuations in the separation of the particles until one (or more) reaches
a breaking point. Many studies considered this problem under one guise or another, mainly with the aim of offering a theory for the {\em mean} breaking time, or the mean rate of breaking $\langle \tau^{-1} \rangle$ where the angular brackets represent an average over many realizations \cite{57Bue,74ZK,90HTB,94DT,94OT,02PS,06SDG,20BW,15LBH}. It turns out (and see below) that the {\em distribution} of breaking 
rates is not at all sharply peaked, and one should worry about the tail of the distribution that represents rare, but potentially catastrophic, fast rates
of breaking (or, mutatis mutandi, short times for failure). We are mainly interested in the probability distribution function (PDF) of breaking rates $P(\tau^{-1};N_b,T,f)$ as a function of $N_b$, $T$ and $f$. We will show that even in this relatively simple problem there exist relatively high probabilities for rupture at times much shorter
than the mean time. In fact, the main result of the Letter is that the PDF of rupture rates exhibits a power law tail, 
\begin{equation}
P(\tau^{-1};N_b,T,f) \sim [\tau^{-1}]^{-\zeta} \ , \quad \text{for}~\tau^{-1}\gg \langle \tau^{-1} \rangle \ ,  
\end{equation}
with a universal exponent $\zeta=2$ (up to higher order terms) {\em independent} of the values of $N_b, T$ and $f$. We here show how to calculate $\zeta$ theoretically and demonstrate excellent agreement with numerical simulations.

\begin{figure}
	\includegraphics[width=0.38\textwidth]{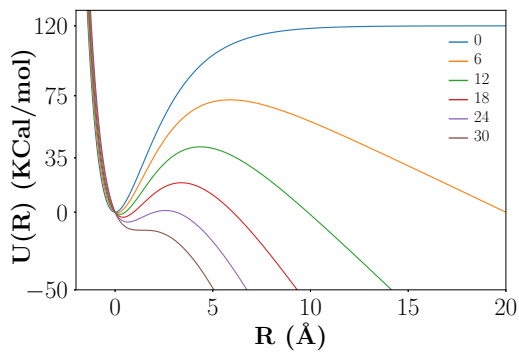}
	\caption{The strained Morse potential with various values of pulling force $f$, see Eq.~(\ref{fmorse}). With the
	present parameters $f_{\rm max} =30$ Kcal/(mole $\times$ \AA); Any force larger than $f_{\rm max}$ 
will result in instant rupture of a bond. Note that $R_{\rm eq}$ increases with $f$ while $R_{\rm b}$ decreases
until they coalesce at $f_{\rm max}$. }
	\label{morse}
\end{figure}

{\bf Model}: The positions of $N_b +1$ beads can be specified by the degrees of freedom ${\bf r}_1,{\bf r}_2, \cdots,{\bf r}_{N_b+1}$ where $\B r\equiv (x,y)$. The $N_b$ bond stretches are denoted $R_1, R_2,\cdots, R_{N_b}$, where $R_i = |{\bf r}_{i+1}-{\bf r}_i | - r_e$ with $r_e=1.54$\AA~being the equilibrium distance between the unstretched bonds. The beads interact via strained Morse potentials of the form \cite{02PS}
\begin{equation}
\label{fmorse}
V(R,f) = D_e (1-e^{-\alpha R})^2 - f R  \ .
\end{equation}
Thus the potential is fully specified by three parameters, $D_e$, $\alpha$ and $f$. In our simulations we used $D_e=120$ Kcal/mole which is the maximum potential energy of the unstretched Morse potential, $\alpha=0.5 $\AA$^{-1}$ is an inverse length scale for reaching this maximal stretch, and $f$ is the applied force in Kcal/(mole $\times$ \AA). In this Letter, these parameters are the same for every bond, but a richer model can be defined with a distribution of parameters.
In a strained Morse potential $V(R,f)$ (see Fig.~\ref{morse}), a potential barrier of size $\Delta(f)$ appears at a distance $R=R_b(f)$. We assume that the chain ruptures if any of the bonds reaches $R=R_b(f)$; no healing phenomena occur in which a chain can reform once it is stretched beyond the peak at $R_b(f)$. 
The barrier height $\Delta(f)$, the positions of the peak  $R_b(f)$ and the minimum of the stretched potential $R_{\rm eq}$ can be calculated as follows
\begin{eqnarray}
\label{bd}
&& \partial V(R,f)/\partial R |_{R = R_b(f)}  =  0 \nonumber \\ 
&& \Delta (f)  =  V(R_b(f),f) - V(R_{eq}(f),f) 
\end{eqnarray}
Plugging Eq.~(\ref{fmorse}) into Eqs.~(\ref{bd}) then yields analytic forms for $R_b(f)$ and $R_{eq}(f)$
\begin{eqnarray}
\label{bd2}
&& R_b(f) = -\frac{1}{\alpha} \log{\frac{1-\sqrt{1-f/f_{max}}}{2} }\nonumber \\ 
&& R_{eq}(f) = -\frac{1}{\alpha} \log{\frac{1+\sqrt{1-f/f_{max}}}{2} }
\end{eqnarray}
where $f_{max} = D_e \alpha/2$.
Below we discuss simulations in which $f$ is not too close to  $f_{\rm max}$. These are the statistically more interesting situations in which the rupture rates $\tau^{-1}$ are widely distributed. For forces too close to $f_{max}$ rupture occurs almost instantly and the problem is less challenging. Note that in general $\tau^{-1}$ is increasing with $f$, becoming singular for $f_{\rm max}$.

For a single Morse potential one can quote the results of Transition State Theory which treats the bond breakage as a unimolecular reaction $R_{eq}\rightarrow R_b$. The {\em mean} rate for single bond breakage has an  Arrhenius form
\begin{equation}
\label{refa}
\langle \tau^{-1}\rangle (R_{eq}\rightarrow R_b,T,f) = \nu(f) \exp{-\Delta(f)/T },
\end{equation}
where the ``attempt rate" to cross the barrier is given by \cite{02PS}
\begin{equation}
\label{refb}
\nu (f) =  \frac{1}{2 \pi} \frac{ \omega_b(f)\omega_{eq}(f)} {\sqrt{\omega_{eq}(f)^2 + 2 \omega_b(f)^2}}  ,
\end{equation}
with 
\begin{eqnarray}
\label{omega}
m \omega_{eq}(f)^2 & = & d^2V(R_{eq},f)/dR_{eq}^2 \nonumber \\
-m \omega_b(f)^2 & =  & d^2V(R_b,f)/dR_b^2 \ .
 \end{eqnarray}
We will see below that the mean rupture rate, for a single bond or the whole polymer, is not sufficiently informative, since the PDF of the polymer rupture rates
is very broad with a power-law asymptotic tail. As said, our aim in this Letter is to go beyond Eq.~(\ref{refa}) to compute the tail of the PDF $P(\tau^{-1};N_b,T,f)$.

\begin{figure}[h!]
	\centering
	\includegraphics[width=0.50\textwidth]{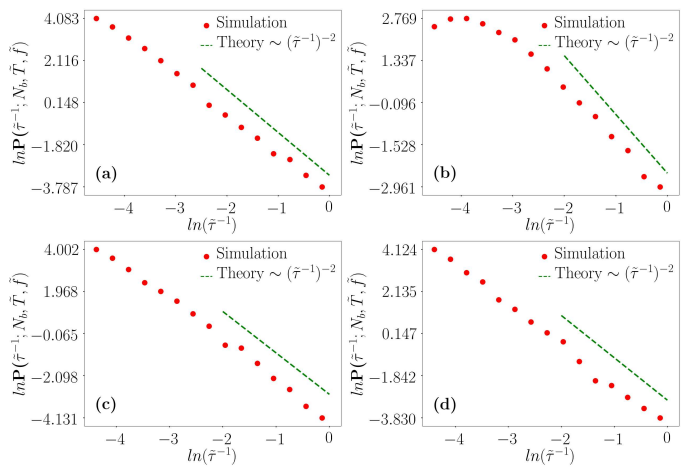}
	\caption{Comparison of the numerically simulated PDF $ P(\tilde \tau^{-1};N_b,\tilde T,\tilde f) $ to the theoretical prediction of its tail,
		for four different sets of parameters. 
		 Here $\tilde \tau^{-1}$ is $\tau^{-1}$ normalized by its maximum value as found in each case. 
	In dots are the simulations and in dashed curves are the theoretical prediction Eq.~(\ref{universal}). Panel a: $N_b=11$, $\tilde T=0.0036$, $\tilde f=0.85$. 
Panel b: $N_b=11$, $\tilde T=0.0146$, $\tilde f=0.67$. Panel c: $N_b=21$, $\tilde T=0.0073$, $\tilde f=0.75$. Panel d: $N_b=21$, $\tilde T=0.0146$, $\tilde f=0.60$.}	\label{compare}
\end{figure}

{\bf Numerical Simulations}: To study the rupture of the polymer we performed molecular dynamics 
simulations employing LAMMPS \cite{Plimpton1995}.  The simulations always begin with the chain of 
$N_b+1$ particles connected sequentially, with the first particle being anchored to a wall at 
position $\B r_1$. Initially, the chain is thermalized by Langevin dynamics at temperature $T$, 
resulting in the chain being folded to a minimum free-energy state. Secondly, this folded 
thermalized polymer is then pulled from the last bead by the force $f$ at a constant temperature. 
Temperature in our simulations is measured in Kelvin, but results are reported in units of $De/k_B$ 
where $k_B =  0.0019872041$ kcal/(mol.K); this rescaled temperature will be denoted as $\tilde T$. 
After thermalization, a constant tensile force $f$ measured in 
units of $f_{\rm max}$ (denoted as $\tilde f$) is applied to the last particle at $\B r_{N_b+1}$. The moment of application of the constant force
is declared to be $t=0$. The simulations are employed to determine the time $\tau$ at which
the chain breaks. For every ensemble we normalize
the rupture times by the maximal inverse time $\tau^{-1}_{\rm max}$ for that ensemble, creating a dimensionless quantity
$\tilde \tau^{-1}\equiv \tau^{-1}/\tau^{-1}_{\rm max}$. 
A typical simulations for a force $\tilde f \approx 0.7$ of a chain with 41 beads at $\tilde T=0.0109$ is shown in the movie that can be observed in the supplementary information.
 During simulations we prepare typically
4000 independent realizations of the polymer, and determine for each the rupture inverse time $\tau^{-1}$. Typical PDF's of $\tilde \tau^{-1}$ as obtained in simulations are shown
	in Fig.~\ref{compare}. The data are shown in a log-log plot to stress that the distribution is very wide. There is high probability to rupture quickly, much quicker than the average rate; the tail of the PDF  
	decays as slowly as a power-law. Our aim is now to understand and compute
	the power-law tail of these distributions. 

{\bf Theory}: Here we present a theory to estimate the PDF $P(\tau^{-1};N_b,T,f)$. We begin with some
elementary statistical mechanics. Define the bond partition function as:
\begin{equation}
\label{part3}
Z_{bond}(T,f) = \int_0^{R_b(f)}dR \exp{[-V(R,f)/T]} .
\end{equation}
We will assume that the moment that any bond reaches $R_b$ the chain breaks instantly and it does not heal. 
Denote now the cumulative probability $C(R,T,f)$ for a single bond to be stretched any distance $0<R<R_b$ . This  is given by:
\begin{equation}
C(R,T,f) = \int_0^R dr \exp{[-V(r,f)/T]}/Z_{bond}(T,f) \ .
\label{prob}
\end{equation}

\begin{figure}
	\includegraphics[width=0.38\textwidth]{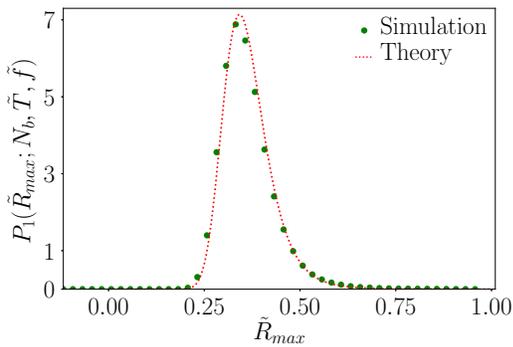}
	\caption{Comparison between the theoretical PDF for maximal bond extension Eq.~(\ref{P1}) and its numerical measurement. Here $N_b=11$, $\tilde T=0.0072856...$ and $\tilde f= 0.77$. $\tilde R_{\rm max}= R_{\rm max}/R_b$.}
	\label{P1com}
	\end{figure}

To connect this to the rate of rupture, we will assert that the bond that breaks is always the bond that has reached
the largest extension among all the bonds, denoted below as $R_{\rm max}$. Obviously, such a bond exists in every realization.
It does not always break, but when rupture occurs, it is always due to the breaking of that bond that was maximally extended. 
Here, we are interested in the situations for which $f< f_{\rm max}$, meaning that the chain reaches thermal equilibrium and equipartition, much before it ruptures. Denote then the probability that {\em any} single bond has a length less than $R$ as $P_<(R)$ and that it is greater than $R$ as $P_>(R)$. 
Using Eq.~(\ref{prob}) we can write
\begin{equation}
P_<(R) =C(R)\ , \quad P_>(R)=1-C(R) \ .
\end{equation}
For $N_b$ bonds, the probability density that exactly one bond will have length larger than $R_{max}$ is:
\begin{eqnarray}
\label{P1}
&&P_1(R_{\rm max};N_b,T,f) = \frac{N_b}{Z_1(T,f)} P_>(R_{\rm max}) [P_<(R_{\rm max})]^{N_b-1} \ , \nonumber\\
&&Z_1(T,f) = N_b \int_0^{R_b(f) }d R P_>(R) [P_<(R)]^{N_b-1} \ .
\end{eqnarray}
Clearly, this result is exact provided that the polymer is thermalized before it ruptures. A typical comparison
to numerical simulations is shown in Fig.~\ref{P1com}. Not surprisingly, agreement is excellent. Note in passing that
for $N_b\to \infty$ this PDF is expected to converge to one of the canonical functions in extreme value statistics, known as the Weibull distribution \cite{01CBTD}.
Obviously it is an analytic function with an end point at $R_{\rm max}=R_b$ where it vanishes. As such it can be expanded
near its end point, 
\begin{equation}
P_1(R_{\rm max};N_b,T,f) =\alpha (R_b-R_{\rm max}) \ , \quad \text{for}~ R_{\rm max}\to R_b \ ,
\label{P1asymp} 
\end{equation}
with $\alpha$ being the derivative at $R_B$ (with dimension of inverse length).

Denote now the breaking rate that we measure in simulations $\tau^{-1}(R_{eq} \rightarrow R_b;N_b,T,f)$. This is 
a random variable, and we will be unable to determine analytically its full distribution. Rather, we will aim at the distribution
of large rates which is dominated by the probability of bonds to stretch to large distances at short times. The quantity of theoretical interest 
will be therefore $\tau^{-1} (R_{\rm max} \rightarrow R_b;N_b, T ,f)$. The PDF of our wanted
rates for rupture is:
\begin{eqnarray}
\label{final}
&&P(\tau^{-1};N_b,T,f) = \int_0^{R_b} dR_{\rm max} \\ &&\times P_1(R_{\rm max};N_b,T,f) \delta(\tau^{-1}-\tau^{-1} (R_{\rm max} \rightarrow R_b; N_b,T ,f)) \ , \nonumber
\end{eqnarray}

Note that the integral is over bond lengths, whereas the delta-function is over rates. To perform the integral we need
to relate the two. To do this, we recognize that the rate $\tau^{-1}$ which appears in the PDF and in the delta function 
is arbitrary, and it will be associated with {\em some} $R_{\rm max}$, say $R_{\rm max}=R^*$ which satisfies the condition
\begin{equation}
\tau^{-1} (R^* \rightarrow R_b;N_b, T ,f) =\tau^{-1} \ .
\label{relate}
\end{equation}
Imagine that we succeeded
to determine the relationship $R^* = R^*(\tau^{-1})$. Then, changing variables accordingly, Eq.~(\ref{final}) leads to 
\begin{equation}
P(\tau^{-1};N_b,T,f)=\frac{dR^*}{d\tau^{-1}} P_1(R^*;N_b,T,f) \ ,
\label{pitau}
\end{equation}
So our task now is to find the Jacobian of the transformation $dR^*/d\tau^{-1}$. 

To find this Jacobian we need to make approximations. 
First, we assume that the total time for breaking can be written as the sum of the time for reaching $R^*$
during the thermal agitation, and then the time from  $R^*$ to $R_b$: 
\begin{equation}
\tau (R_{eq} \rightarrow R^*) + \tau (R^* \rightarrow R_b)  = \tau (R_{eq} \rightarrow R_b)  \ .
\label{sum0}
\end{equation}
Since we are interested in the rates, we invert this equation in favor of the fast rate going from $R^*$ to $R_b$:
\begin{equation}
\tau^{-1} (R^*\rightarrow R_b) = [\tau (R_{eq} \rightarrow R_b)-\tau (R_{eq} \rightarrow R^*)]^{-1} \ .
\label{suminv}
\end{equation}
Second, we assume that the major source of randomness is in the distribution $P_1(R_{max})$ which is governed by the thermal
agitation. Therefore, {\em for the purpose of estimating the Jacobian}, the random times appearing in Eq.~(\ref{suminv}) can be estimated by their means. Thus
\begin{eqnarray}
\label{taurmax2}
&&\tau^{-1}(R_{eq} \rightarrow R_b; N_b, T ,f) \sim \nu(f) e^{-\Delta (f)/T}  \\
&& \tau^{-1}(R_{eq} \rightarrow R^*;N_b, T ,f) \sim \nu(f) e^{[V(R_{eq},f) -V(R^*,f)]/T}  \  . \nonumber
\end{eqnarray}

Returning now to Eq.~(\ref{suminv}), we note that if $R^*$ is close to $R_b$ where the potential has a maximum, 
we can estimate to second order in $R_b-R^*$,
\begin{equation}
V(R_b,f)-V(R^*,f)=\frac{m\omega^2_b(f)}{2}(R_b-R^*)^2 \ .
\label{expansion}
\end{equation}
Using now Eqs.~(\ref{taurmax2}) and (\ref{expansion}) in Eq.~(\ref{suminv}) results in
\begin{equation}
R_b-R^* =\sqrt{\frac{2T}{m\omega_b^2(f)}}\left[-\ln[1-\frac{\nu(f) \exp(-\Delta (f)/T)}{\tau^{-1}}]\right]^{1/2} \ .
\label{invert}
\end{equation}	
Computing the derivative of $R^*$ with respect $\tau^{-1}$ we end up with 
\begin{eqnarray}
&&\frac{dR^*}{d\tau^{-1}} =  \sqrt{\frac{T}{2m\omega_b^2(f)}}\left[-\ln[1-\frac{\nu(f) \exp(-\Delta (f)/T)}{\tau^{-1}}]\right]^{-1/2}\nonumber\\
&&\frac{\nu(f) \exp(-\Delta (f)/T)}{\tau^{-1}[\tau^{-1} -\nu(f) \exp(-\Delta (f)/T)]} \ . \label{finalder} 
\end{eqnarray}	
 Eqs.~(\ref{pitau}) and (\ref{finalder}) are our theoretical predictions that should be compared with the results of numerical simulations. 
 For $\tau^{-1}\gg \langle \tau^{-1} \rangle$ we find that the Jacobian Eq.~(\ref{finalder} ) goes as
 \begin{equation}
\frac{dR^*}{d\tau^{-1}} \propto [\tau^{-1}]^{-3/2} \ , \quad \text{for}~\tau^{-1}\gg \langle \tau^{-1} \rangle  \ ,
 \end{equation}
 up to higher order terms. 	
On the other hand, combining Eqs.~(\ref{P1asymp}) and (\ref{invert}) we find that 
\begin{equation}
P_1(R^*;N_b,T,f)\sim [\tau^{-1}]^{-1/2} \ ,
\end{equation}
again up to higher order terms. 
 Together, these two factors result in a power law tail for  $P(\tau^{-1};N_b,T,f) $,
 \begin{equation}
 P(\tau^{-1};N_b,T,f) \sim  [\tau^{-1}]^{-2} \ , \quad \text{for}~\tau^{-1}\gg \langle \tau^{-1} \rangle \ ,
 \label{universal}
 \end{equation}
 up to higher order corrections.  We note that this result is independent of $N_b$, $T$, and $f$ as long as
 the polymer equilibrates before snapping. This is a surprisingly universal result that needs to be compared to simulations. 

{\bf Comparison of theory and simulations}: In comparing our theory to the numerical simulations, we remember that the former is limited to the tail of the PDF 
$ P( \tau^{-1};N_b,T,f)$ for $\tau^{-1}\gg \langle \tau^{-1} \rangle $. The actual distribution is normalized but the theoretical prediction is not. The comparison is presented in Fig.~\ref{compare} for four different choices
of parameters. In particular we note that the prediction of a power law tail is very well supported, and 
the exponent in the power law tail computed theoretically fits well the tail of the actual normalized
distribution as found in the simulations. The agreement between theory and simulations shown in Fig.~\ref{compare}
is typical {\em as long as the polymer has reached thermal equilibrium before rupture}. We have also considered forces $f$ that are too high for
the polymer to equilibrate (not shown here), and not surprisingly the excellent agreement exhibited in Fig.~\ref{compare} disappears. In such cases,
also the comparison of the measured $P_1(R_{\rm max};N_b,T,f)$ to the theoretical result Eq.~(\ref{P1}) as shown in Fig.~\ref{P1com} is no longer favorable.

{\bf Summary and Discussion}: In summary, we show that the ability to predict the tail of the PDF for fracture rates
in the strained polymer chains, and also in many other similar problems, depends on two ingredients. The first is an identification of
the ``weakest link", which in this case is the bond that extends most, denoted above as $R_{\rm max}$. A first step
of the analysis requires a calculations of the PDF of this weakest link. When the random perturbations are thermal,
standard statistical mechanics suffices to compute the PDF. If the random perturbations are of a different sort,
their nature and their statistics must be provided in order to achieve this first step. The second step is where
our approach appears novel, in determining the rate of failure associated with each value of the maximally dangerous
link. In the present example, it is Eq.~(\ref{invert})  that provided the necessary relation. In any other problem 
of a similar type, physical intuition should be exercised again to state the analogous relation. Only the combination
of these two steps can provide predictability of the type shown in Fig.\ref{compare}. In future work. one will need
to explore these ideas in more complex models like bundles of polymers, say of poly-ethilene oxide (PEO) \cite{20BW}, protein gels \cite{14LPDM}
and other biological examples, cf. Refs.~\cite{00Sei,10HCB}.

\acknowledgments We thank Eivind Bering and Astrid de Wijn for useful discussions that gave birth to this project.
We are indebted to Didier Sornette for very useful comments on a draft of this Letter. 
This work was supported in part by the Minerva Center for ``Aging, from Physical Materials to Human Tissues", the scientific and cooperation agreement between Italy and Israel through the project COMPAMP/DISORDER,
and the Research Council of Norway through its Centres of Excellence funding scheme, project number 262644

\bibliography{ALL}

\end{document}